\documentclass[aps,prl,reprint,twocolumn,superscriptaddress,showpacs]{revtex4-1}

\usepackage[utf8]{inputenc}
\usepackage[english]{babel}
\usepackage{float}
\usepackage{amsfonts,amsmath,amssymb}
\usepackage{bm}
\usepackage{graphicx}
\usepackage{color}
\usepackage{array,multirow,makecell}

\graphicspath{{}}

\newcommand{\bz}{\bm{z}}

\newcommand{\bE}{\bm{E}}
\newcommand{\bV}{\bm{V}}
\newcommand{\bpsi}{\pmb{\psi}}

\newcommand{\TE}{\mathrm{TE}}
\newcommand{\TM}{\mathrm{TM}}
\newcommand{\dB}{\mathrm{dB}}

\begin{document}

\title{Quantum reflection of antihydrogen from a liquid helium film}

\author{P.-P. Cr\'epin} \email[]{pierre-philippe.crepin@lkb.upmc.fr}
\affiliation{Laboratoire Kastler Brossel (LKB), UPMC-Sorbonne
Universit\'es, CNRS, ENS-PSL Research University, Coll\`ege de
France, 75252, Paris, France}
\author{E.A. Kupriyanova}
\affiliation{P.N. Lebedev Physical Institute, 53 Leninsky prospect,
117924, Moscow, Russia}
\affiliation{Russian Quantum Center, 100 A, Novaya street, Skolkovo, 
143025, Moscow, Russia}
\author{R. Gu\'erout} \email[]{romain.guerout@lkb.upmc.fr}
\affiliation{Laboratoire Kastler Brossel (LKB), UPMC-Sorbonne
Universit\'es, CNRS, ENS-PSL Research University, Coll\`ege de
France, 75252, Paris, France}
\author{A. Lambrecht} 
\affiliation{Laboratoire Kastler Brossel (LKB), UPMC-Sorbonne
Universit\'es, CNRS, ENS-PSL Research University, Coll\`ege de
France, 75252, Paris, France}
\author{V.V. Nesvizhevsky}
\affiliation{Institut Laue-Langevin (ILL), 71 avenue des Martyrs,
 38042, Grenoble, France}
\author{S. Reynaud} \email[]{serge.reynaud@lkb.upmc.fr}
\affiliation{Laboratoire Kastler Brossel (LKB), UPMC-Sorbonne
Universit\'es, CNRS, ENS-PSL Research University, Coll\`ege de
France, 75252, Paris, France}
\author{S. Vasiliev}
\affiliation{Department of Physics and Astronomy, 
University of Turku, 20014, Turku, Finland}
\author{A.Yu. Voronin}
\affiliation{P.N. Lebedev Physical Institute, 53 Leninsky prospect,
117924, Moscow, Russia}
\affiliation{Russian Quantum Center, 100 A, Novaya street, Skolkovo, 
143025, Moscow, Russia}

\date{\today}

\begin{abstract}
We study the quantum reflection of ultracold antihydrogen atoms bouncing on the surface of a liquid helium film. The Casimir-Polder potential and quantum reflection are calculated for different thicknesses of the film supported by different substrates. Antihydrogen can be protected from annihilation for as long as 1.3s on a bulk of liquid $^4$He, and 1.7s for liquid $^3$He. These large lifetimes open interesting perspectives for spectroscopic measurements of the free fall acceleration of antihydrogen. Variation of the scattering length with the thickness of a film of helium shows interferences which we interpret through a Liouville transformation of the quantum reflection problem.
\end{abstract}

\maketitle

\section{Introduction}
Quantum reflection is a non classical phenomenon which appears when a quantum matter wave approaches 
a rapidly varying attractive potential. Instead of accelerating towards the surface, the quantum particle has a 
probability to be reflected. This process has been studied theoretically for the van der Waals potential since 
the early days of quantum mechanics \cite{Lennard-Jones1936III,Lennard-Jones1936IV,Berry1972,Friedrich2004a}. 
It was first observed experimentally for H and He atoms \cite{Nayak1983,Berkhout1989,Yu1993} and then for ultracold atoms or molecules on solid surfaces \cite{Shimizu2001,Druzhinina2003,Pasquini2004,Pasquini2006}. 

In the last years quantum reflection has been studied also for antimatter \cite{Voronin2005pra,Voronin2011,Voronin2012pra} since it should play a key role in experiments with antihydrogen atoms  \cite{Dufour2013qrefl,Dufour2013porous,Dufour2014shaper}. It was shown that the free fall acceleration of antihydrogen can in principle be evaluated accurately \cite{Crepin2017} through spectroscopic studies of the quantum levitational states \cite{Breit1928,Nesvizhevsky2002nature} of atoms trapped by quantum reflection and gravity \cite{Jurisch2006,Madronero2007}. Following uncertainty principle of quantum mechanics, such spectroscopic measurements should have a better accuracy for larger lifetime of antihydrogen in the trap. 

In this letter, we calculate quantum reflection of antihydrogen above liquid helium films and show that the lifetime 
of antihydrogen reaches values as high as 1.3s for a bulk (a thick film) of liquid $^4$He and 1.7s for a bulk of 
liquid $^3$He. We also study the effect of thickness for $^4$He films supported by different substrates and bring out a
surprising interference pattern for the scattering length as a function of thickness. By using a Liouville transformation 
of the quantum reflection problem, we propose an interpretation of this phenomenon in terms of shape resonances. 

\section{Casimir-Polder potential}
\label{sec:CP}

\begin{figure}[ht]
   \center
   \includegraphics[scale=0.5]{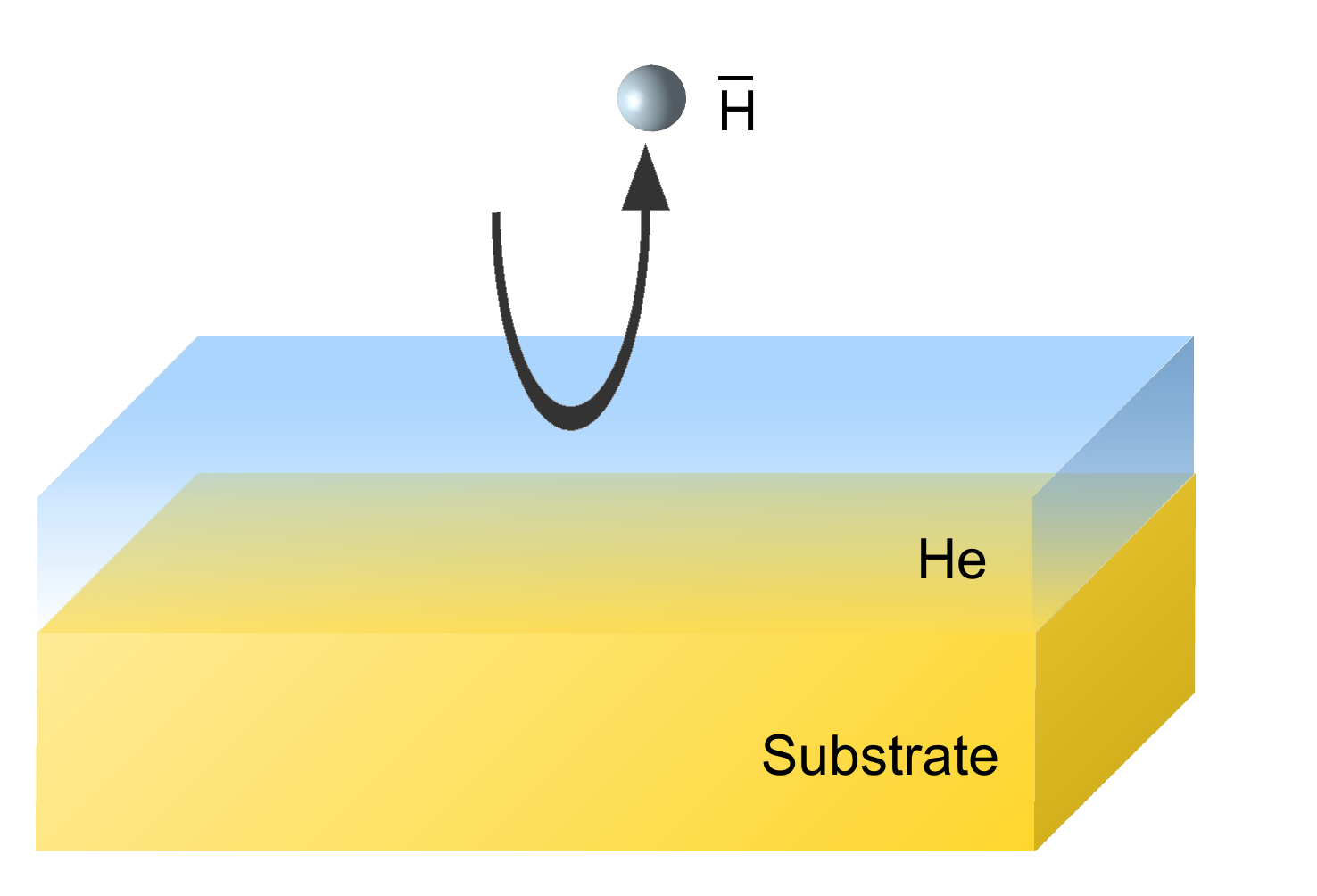}
   \caption{Representation of the quantum reflection process for an antihydrogen atom falling onto a helium film supported by a substrate. We study the limiting case of a bulk of helium (very large thickness of the film) as well as the general case of a film of finte thickness $d$ supported by a substrate. }
   \label{schema}
\end{figure}

We study quantum reflection for an antihydrogen atom of mass $m$ falling onto a liquid helium film of thickness $d$ supported by a substrate (see Fig.~\ref{schema}). 

The atom is sensitive to the Casimir-Polder (CP) potential and to the free fall acceleration $\overline{g}$. 
In the present letter, we focus our attention on the quantum reflection in the CP potential 
\footnote{The full problem of combined effects of the CP and gravity potentials is treated in \cite{Crepin2017}.}, 
as the latter is effective at distances $z$ much smaller than the length scale $\ell_g$ associated with quantum effects 
in the gravity field  $\ell_g=(\hbar^2/2m^2g)^{1/3}\simeq 5.87\,\mu\text{m}$. Note that we suppose 
$\overline{g}=g$ in all numerical evaluations. 

The CP potential is evaluated at zero temperature as the following function of the distance $z$ of the atom 
to the liquid helium surface \cite{Messina2009}
\begin{eqnarray}
&&V(z)=\frac{\hbar}{c^2}\int_0^\infty\frac{d\xi}{2\pi}\xi^2 \alpha(i\xi)\int \frac{d^2\mathbf{q}}{(2\pi)^2} \frac{e^{-2\kappa z}}{2\kappa} \sum_p s_p r_p  ~, \\
&&\kappa = \sqrt{\xi^2/c^2+\mathbf{q}^2}\;,\quad
s_\TE=1\;,\quad 
s_\TM=-\frac{\xi^2+2c^2\mathbf{q}^2}{\xi^2} ~.\notag
 \label{pot}
\end{eqnarray}
This expression is integrated over the complex frequency $\omega=i\xi$ and
the transverse wave-vector $\mathbf{q}$, and 
summed up over the two field polarizations $p=\TE,\TM$.  

The dynamic polarizability $\alpha(\omega)$ of the antihydrogen atom is
supposed to be the same as for the hydrogen atom, because possible differences 
between the two cases would be too small to have an influence at the level of precision
aimed in the present study. 
Note that there are differences between hydrogen and antihydrogen for the atomic physics 
in the proximity of the helium film.
While the potential of interaction for a hydrogen atom leads to the existence of a 
single bound state with an adsorption energy of the order of 1K on $^4$He bulk
\cite{Castaing1983,Kagan1984,Berkhout1993}, the repulsive part of potential  may be
absent in the much less studied case of antihydrogen. 
These differences are not studied in this letter where we focus attention on antihydrogen for which
quantum reflection is the only reason for scattering of atoms back from the helium surface.

The reflection amplitudes  $r_p$ are calculated for polarizations $p$ by combining 
the Fresnel amplitudes at interfaces and propagation in the helium film. 
The optical properties of $^4$He are described with a sufficient accuracy by a model dielectric constant 
with three resonances \cite{Sabisky1972} 
\begin{eqnarray}
&&\epsilon(i\xi) \simeq 1 + \sum_{k=1,2,3}\frac{a_k}{1+(\xi/\omega_k)^2} \label{epsilon} ~, \\
&&\left(\omega_1,\omega_2,\omega_3\right)=\left(3.22,3.74,12\right)\times10^{16}\textrm{rad.s}^{-1}
~, \notag \\
&&\left(a_1,a_2,a_3\right)=\left(0.016,0.036,0.0047\right)
~. \notag
\end{eqnarray}
This model corresponds to a dielectric constant close to unity at the static limit ($\epsilon(0)-1\simeq0.0567$) 
as well as at all frequencies. We also use the optical model \eqref{epsilon} for $^3$He, with the same 
resonance frequencies $\omega_k$, and the resonance amplitudes $a_k$ multiplied 
by the same factor calculated to reproduce the static dielectric constant $\epsilon(0)-1\simeq0.043$ 
known from experiments \cite{Kierstead1976}.

These numbers lead to a poor reflectance of the film for electromagnetic waves and weak values 
for the CP potential with values even weaker for $^3$He than for $^4$He.
It follows that quantum reflection occurs closer to the material surface where the CP potential is
much steeper, which explains the large quantum reflection probability found below, 
with reflection even larger for $^3$He than for $^4$He.
In both cases we use an effective dielectric constant and disregard the role played
by excitations in the helium film like ripplons. 
The latter is well justified at temperatures below 100 mK \cite{Goldman1986}, 
the temperature range where results obtained in the following are accurate.

We now discuss the results calculated for the CP potential in different situations. 
We begin with the limiting cases of $d\rightarrow0$ and $d\rightarrow \infty$ where
we obtain respectively the CP potentials of the naked substrate and of liquid helium bulk
(that is liquid helium film with a large thickness).
These potentials, attractive at all distances, behave as non retarded  \textit{van der Waals} potentials 
at short distances  and as \textit{retarded} potentials at large distances,
with the two domains separated by the wavelength $\lambda_A \simeq 121$nm
of the first atomic transition 1S$\rightarrow$2P of antihydrogen
\begin{eqnarray}
&&V(z) \simeq -\frac{C_3}{z^3} \;,\quad z \ll \lambda_A ~, \label{nonretarded} \\
&&V(z) \simeq -\frac{C_4}{z^4} \;,\quad z \gg \lambda_A ~. \label{retarded}
\end{eqnarray} 
Constants $C_3$ and $C_4$ are given in table~\ref{coeff} for liquid helium bulk and substrates 
made of silica, silicon or gold. 

\begin{table}[h]
\centering
\begin{tabular}{|c|c|c|}
\hline  medium  & \,$C_3$  [$\text{E}_\text{h}\text{a}_0^3$] \, & \, $C_4$  [$\text{E}_\text{h}\text{a}_0^4$] \, \\
\hline \hline\, liquid $^3$He  \,      &   0.0034   &  1.19   \\
\hline            liquid $^4$He          &   0.0045   &  1.55   \\
\hline            silica  		           &  0.053    & 28.1     \\
\hline            silicon  		       &  0.101    & 50.28    \\
\hline            gold  		          &  0.085    & 73.38    \\
\hline
\end{tabular}
\caption{Constants $C_3$ and $C_4$ for bulks of liquid helium and substrates 
made of silica, silicon or gold, expressed in atomic units
 ($\textrm{E}_\textrm{h}$ and $\text{a}_0$ are the Hartree energy and Bohr radius).}
\label{coeff}
\end{table}

For the purpose of discussing the influence of the optical properties of the mirror
on the CP potentials, we normalize the potential $V(z)$ obtained from \eqref{pot}
by the potential $V_*(z)=-C_4^*/z^4$  corresponding to the large distance limit
above a perfectly reflecting mirror.  
The constant $C_4^*=-3\alpha(0)\hbar c/(32 \pi^2)$ is determined by the 
static polarizability $\alpha(0)=\tfrac92\text{a}_0^3$ of antihydrogen. 

\begin{figure}[bh]
   \center
   \includegraphics[scale=0.36]{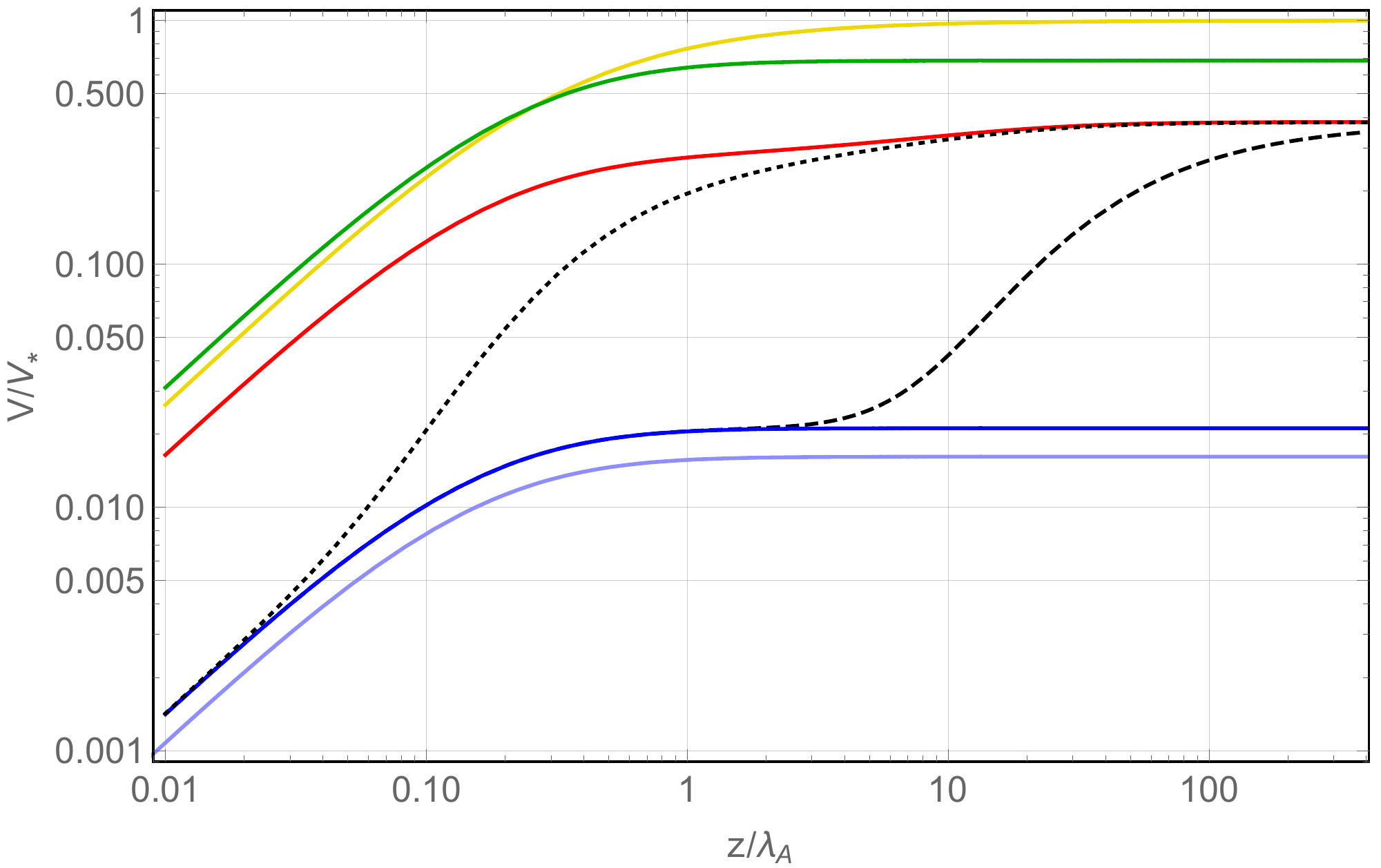}
   \caption{CP potentials $V(z)$ normalized by the potential $V_*(z)$ calculated for a perfect mirror at large distances. Distances $z$ are normalized by the wavelength $\lambda_A \simeq 121$nm of the 1S$\rightarrow$2P antihydrogen
transition.
The full lines correspond, from bottom to top, to bulks of $^3$He (light blue),  $^4$He (dark blue),
silica (red),  silicon (green) and  gold (yellow).  
The other lines correspond to liquid helium films of thickness $d=10\lambda_A$ (dashed line) and 
$d=0.1\lambda_A$ (dotted line) on a silica bulk. }
   \label{potential}
\end{figure}

The ratios $V(z)/V_*(z)$ obtained for liquid $^3$He and $^4$He bulks as well as
silica, silicon and gold bulks are plotted as full lines in Fig.~\ref{potential}. 
The ratios obtained for liquid $^4$He films with finite thickness $d$ on silica 
are plotted in Fig.~\ref{potential} as dashed lines. 
They go smoothly from the one obtained for 
a liquid helium bulk for $z\ll d$ to that for a silica bulk for $z\gg d$. 

\section{Quantum reflection from a liquid helium bulk}
\label{sec:helium}

The previous calculations show a very low value of the CP potential for thick enough liquid helium films,
as explained by the fact that liquid helium is almost transparent for the electromagnetic field. 
We now discuss the consequence of this fact in terms of large quantum reflection from liquid helium bulks.  

\begin{figure}[ht]
   \center
   \includegraphics[scale=0.35]{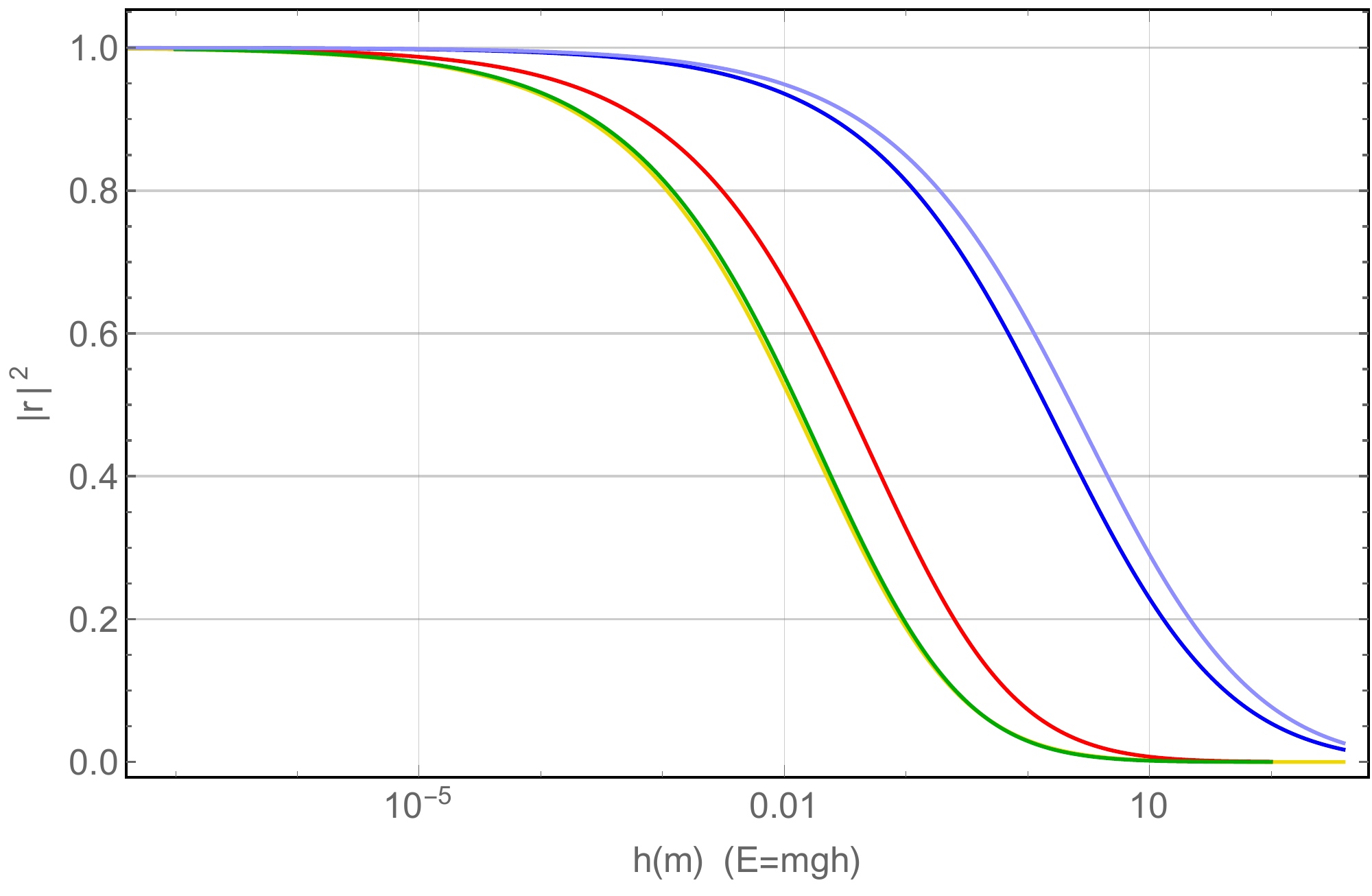}
   \caption{Quantum reflection probability  as a function of the free fall height of the atom $h$, 
that is also of its energy  $E=mgh$. The full lines correspond, from top to bottom, to bulks of 
$^3$He (light blue),  $^4$He (dark blue), silica (red),  silicon (green) and  gold (yellow). } 
   \label{reflectivity}
\end{figure}

To this aim, we solve the Schr\"odinger equation for the antihydrogen falling into the CP 
potential \cite{Dufour2013qrefl} above the liquid helium film. We then obtain the reflection amplitude $r$ 
as the ratio of the outgoing wave to the incoming one far from the film. 
The quantum reflection probability is the squared modulus of this amplitude $R=|r|^2$. The results are shown 
in Fig.~\ref{reflectivity} with larger and larger probability obtained for the weaker and
weaker potentials of Fig.~\ref{pot}. In particular, quantum reflection for atoms falling from a height 
$h$ and thus having a given energy $E=mgh$   is much larger on a liquid helium bulk than on the 
other materials studied here.

We then extract from the reflection amplitude the complex length $\mathcal{A}(k)$ depending
on the wavevector $k$ equivalent to the energy $E=\hbar^2k^2/(2m)$
\begin{equation}
\mathcal{A}(k)\equiv-\frac{i}{k}\frac{1+r(k)}{1-r(k)}.
\end{equation}
The effective range theory modified to account for the $-C_4/z^4$ tail of the 
potential leads to the expression 
\begin{align}
\mathcal{A}(k)&=-i\ell\,\alpha(k\ell)  ~,\\
 \alpha(K)&=\alpha_0+i\frac{\pi}{3}K+\left(\alpha_2+\frac{4}{3}\alpha_0\ln K \right)K^2 ~,  \notag
\end{align}
where $\ell=\sqrt{2mC_4}/\hbar$ is the characteristic length associated with the constant $C_4$ 
while $\alpha_0$ and $\alpha_2$ are two dimensionless parameters. 
The values of $\alpha_0$ and $\alpha_2$ are known for the special case $V=-C_4/z^4$
\cite{Omalley1961,Arnecke2006}. In the problem studied here, the CP potential above the liquid helium film
does not reduce to this special form, and we proceed as in \cite{Crepin2017} by extracting the 
2 parameters from a fit of the low$-k$ dependence of $r(k)$. 

We now discuss the results thus obtained for the \textit{scattering length} $a=-i\ell\,\alpha_0$
which is deduced from the characteristic length $\ell$ and the parameter $\alpha_0$.
For quantum reflection on a liquid $^4$He bulk, one finds for example 
$a=-\left(34.983+44.837i\right)a_0$. The imaginary part $b=-\text{Im}(a)$ of this scattering length
determines the mean lifetime $\tau $ for atoms bouncing above the bulk \cite{Crepin2017}
\begin{equation}
\tau = \frac{\hbar}{2 m g b}   ~.
\end{equation}
In table~\ref{lifetime}, we compare this values obtained for $\tau $ from the quantum reflection 
propabilities drawn on  Fig.~\ref{reflectivity} and also for porous silica studied in \cite{Dufour2013porous}.
We also give the values for the number $N_1$ of bounces for an atom in the first quantum levitation state.
The numbers show that liquid helium is a much better reflector for antihydrogen matter waves
than the other materials which have been studied up to now. The much larger lifetime, that is also the much larger 
number of bounces before annihilation, implies that it should be possible to trap antimatter for long enough 
to improve significantly the spectroscopy measurements discussed in \cite{Crepin2017}.

\begin{table}[h]
\centering
\begin{tabular}{|c|c|c|} 
\hline material  & \,  $\tau$ [s] \, & \, $N_1$ \, \\
\hline \hline  perfectly reflective     &  0.11 &  33\\
\hline 	silica bulk	&  0.22 & 66 \\
\hline \,	porous silica ($95\%$ porosity)\,	 &  0.94  & 282\\
\hline 	liquid $^4He$ bulk	& 1.35 & 405 \\
\hline 	liquid $^3He$ bulk	& 1.71 & \, 514 \, \\
\hline
\end{tabular}
\caption{Lifetime $\tau$ of antihydrogen in seconds above various material surfaces and number $N_1$ of bounces 
for an atom in the first quantum gravitational state for different bulk materials and for porous silica 
(see \cite{Dufour2013porous} for the latter case).}
\label{lifetime}
\end{table}

\section{Quantum reflection from finite thickness films}
\label{sec:substrate}

We now investigate the effect on quantum reflection
 of the finite thickness of a liquid $^4$He film supported by a substrate. 
We present the results of the calculations in terms of the scattering length $a$
which now depends on the thickness $d$ of the film as well as 
on the optical properties of the substrate.

\begin{figure}[th]
   \center
   \includegraphics[scale=0.36]{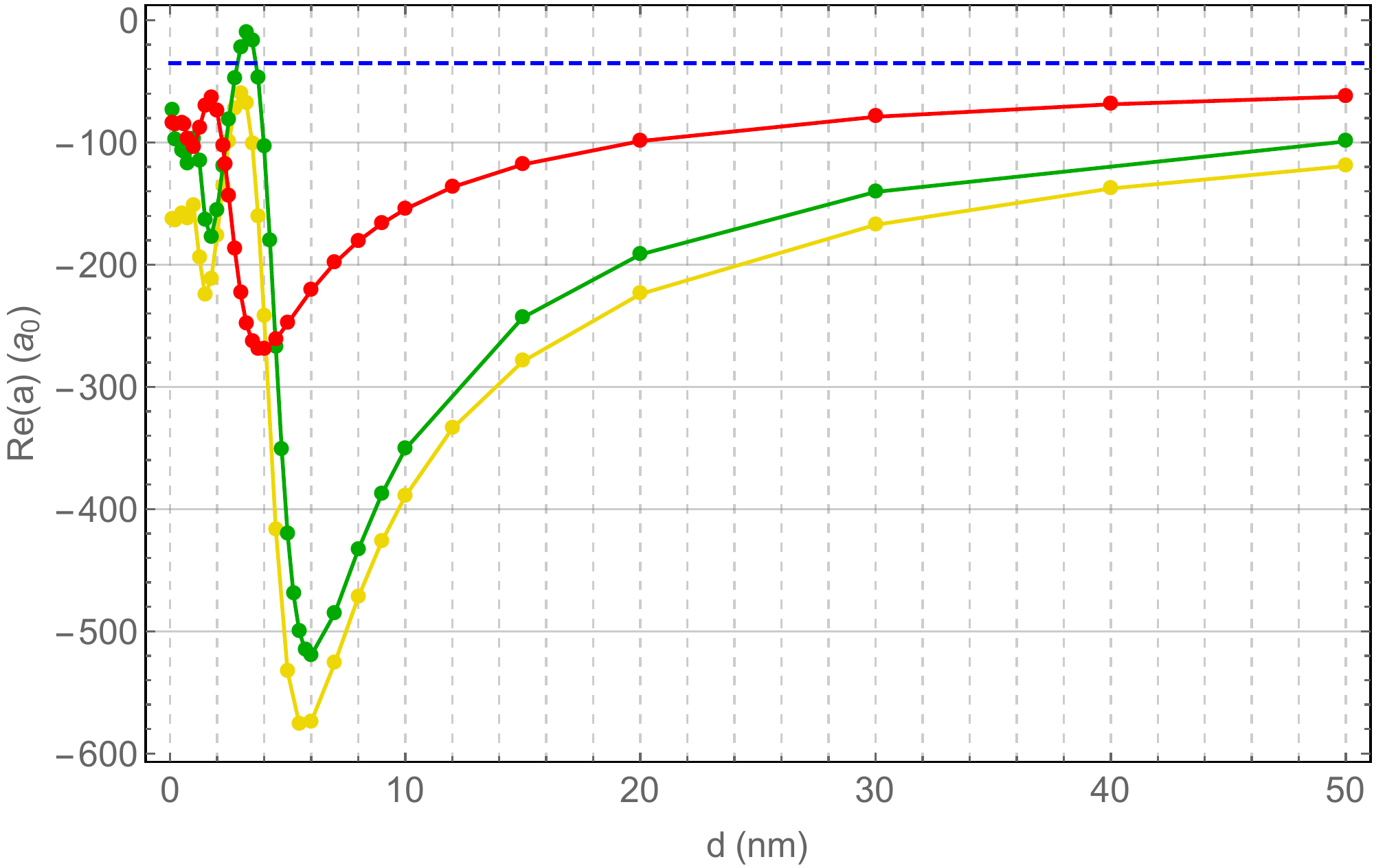}
   \includegraphics[scale=0.36]{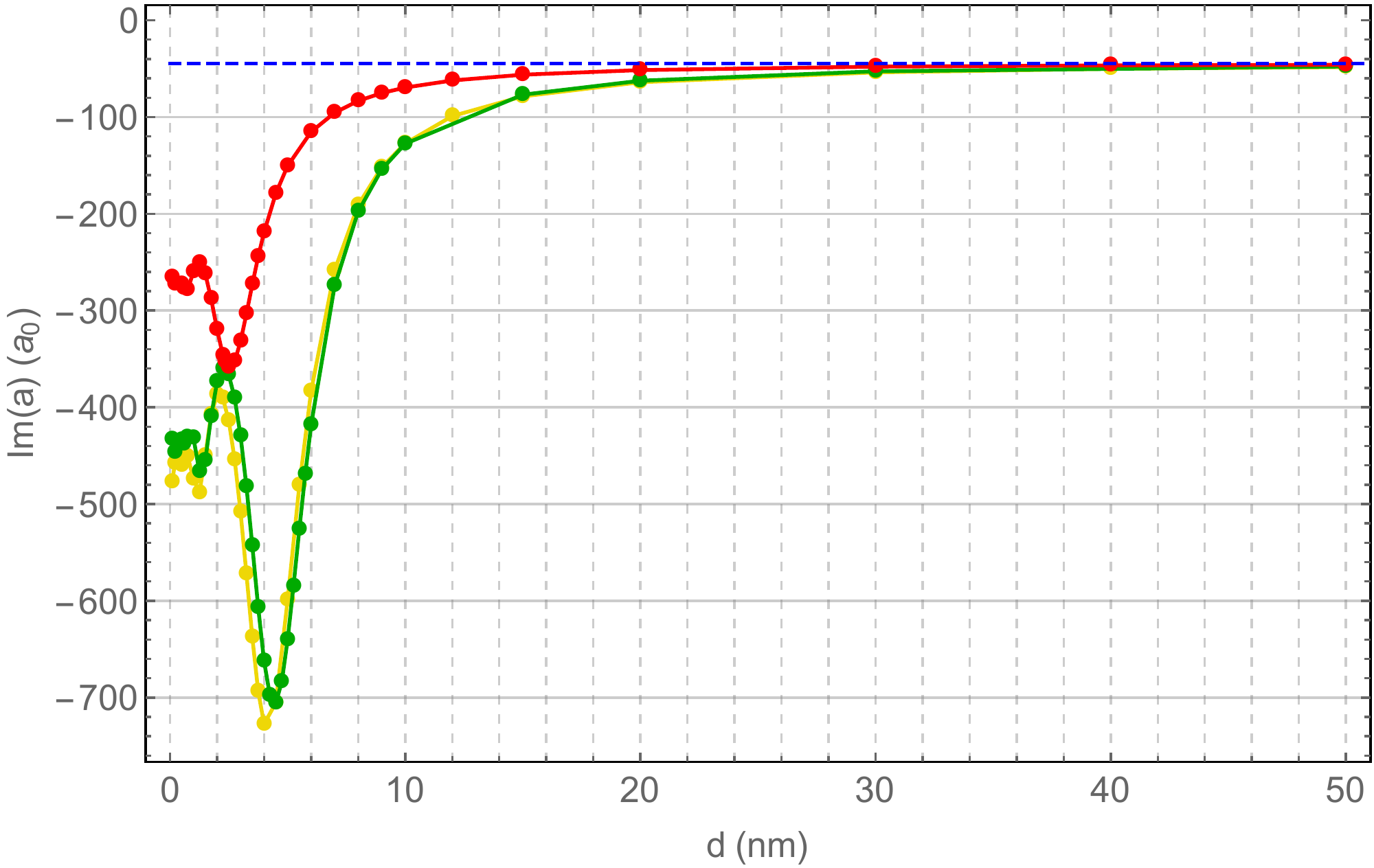}
   \caption{Real (upper plot) and imaginary (lower plot) parts of the scattering length depending 
on the thickness $d$ of the liquid $^4$He film, drawn from the top to the bottom for a silica substrate (red curve),
a silicon substrate (green) and a gold substrate (yellow). For comparison, the dashed (blue) line corresponds to 
real and imaginary parts of the scattering length for a liquid $^4$He bulk.  }
   \label{a}
\end{figure}

\begin{figure}[h]
   \center
   \includegraphics[scale=0.36]{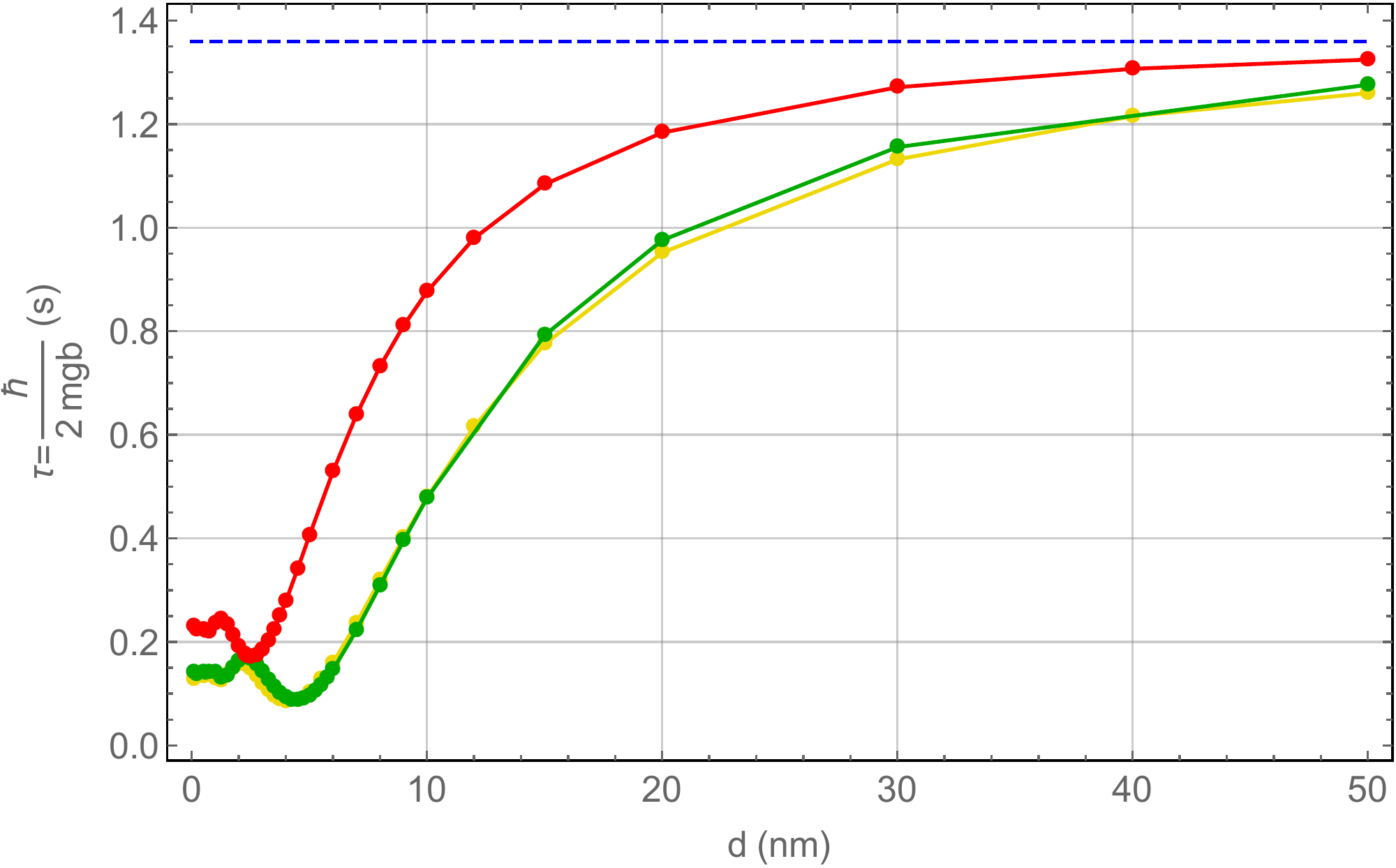}
   \caption{Lifetime $\tau$ depending on the thickness $d$ of the liquid $^4$He film, drawn from the top 
to the bottom for a silica substrate (red curve), a silicon substrate (green) and a gold substrate (yellow). 
For comparison, the dashed (blue) line is the lifetime corresponding to the liquid $^4$He bulk. }
   \label{tau}
\end{figure}

Results are presented in Fig.~\ref{a} for films supported by silica, silicon and gold substrates. 
The important effect of the thickness is clearly seen on this plot. 
For thicknesses larger than a few tens of nanometers, the scattering length 
reaches asymptotically the value found above for a liquid $^4$He bulk. 
The curves give the thickness of the film to be chosen sufficient for recovering the large lifetimes
predicted at the limit of the bulk. 
This property is also illustrated in terms of variation of the lifetime in Fig.~\ref{tau}. 
As could be expected, the substrate which leads to the larger lifetime for a given thickness of the liquid helium film 
is the one which would have the best reflectivity without the film (silica in our case).

\begin{figure}[t]
   \center
   \includegraphics[scale=0.3]{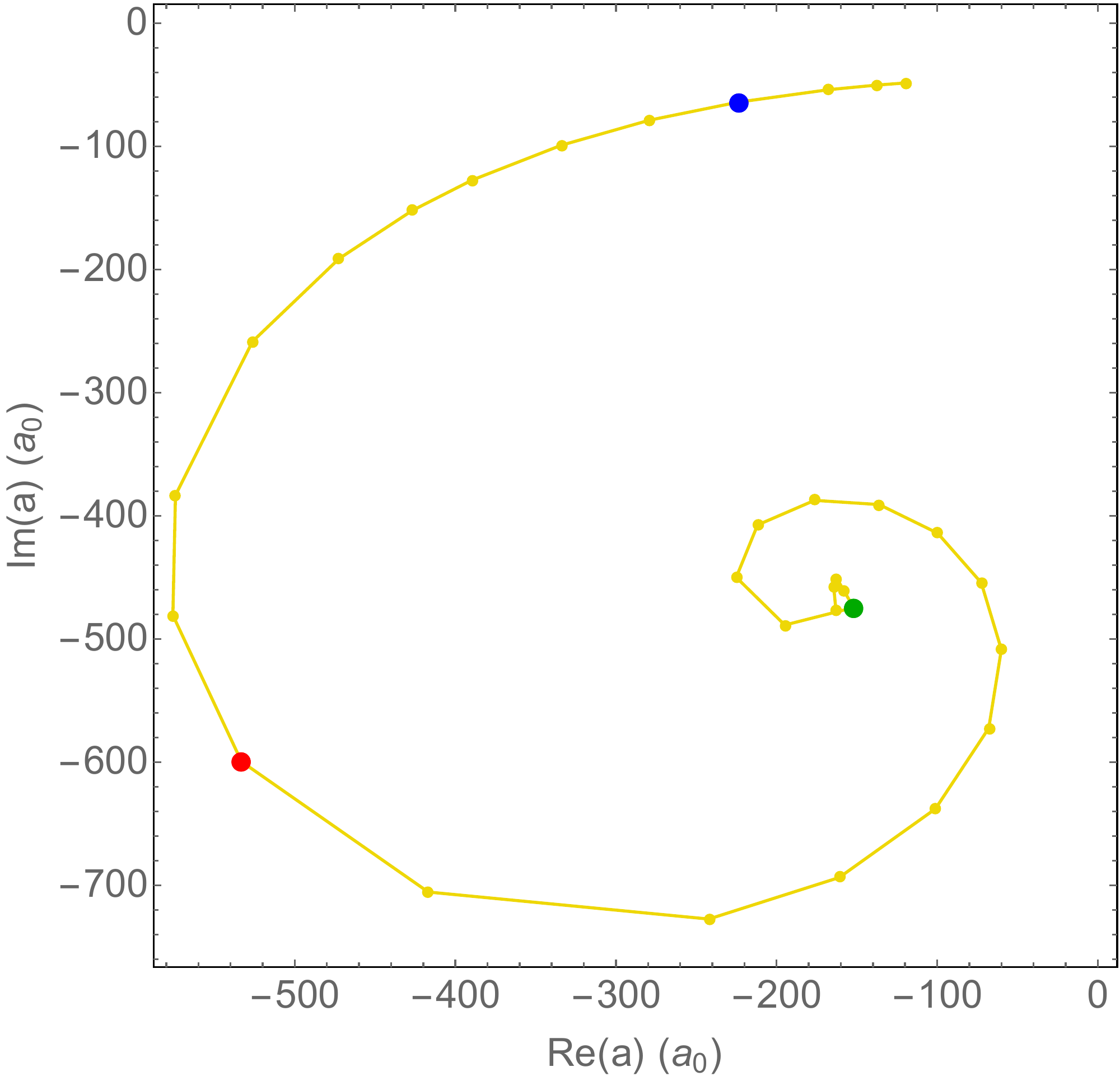}
   \caption{Scattering length $a$ represented in the complex plane, depending on the thickness $d$ of the liquid helium film above a gold substrate. The thickness ranges from 0.1nm (center of the spiral) to 50nm (outer part of the spiral). Green point corresponds to $d=1$nm, red to $d=5$nm and blue to $d=20$nm.}
   \label{quadrature}
\end{figure}

For small thicknesses, of the order of a few nanometers, real and imaginary parts of the scattering length 
are found to oscillate in phase quadrature in Fig.~\ref{a}-\ref{tau}.
This property is confirmed by the variation of $a$ in the complex plane, shown in Fig.~\ref{quadrature}
in the case of a gold substrate.
It looks like a consequence of an interference phenomenon that we discuss in the next section.

\section{Discussion of the oscillations}
\label{sec:liouville}

The counterintuitive property of a larger quantum reflection for a weaker CP potential 
\cite{Dufour2013qrefl,Dufour2013porous} has been discussed in recent papers
by using Liouville transformations of the Schr\"odinger equation. 
Such transformations allow one to map the problem of quantum reflection from an atractive well
 to the more intuitive problem of ordinary reflection from a wall, with the latter becoming higher 
for weaker CP potentials \cite{Dufour2015epl,Dufour2015jpb}. 

The shape of the CP potential for the film of thickness 10$\lambda_A$ in Fig.~\ref{potential} suggests that 
the atom falling onto the film see two zones of rapid variation of the potential, the first one at the transition 
from the potential which would be seen for the naked substrate to that of a helium bulk, 
and the second one at the approach to the liquid helium film. 
Using Liouville transformations,
we now interpret the oscillations of the scattering length seen in Fig.~\ref{quadrature}
as an interference between reflections on the two walls.

Liouville transformations \cite{Liouville1837,Olver1997} are gauge transformations of the Schr\"odinger equation  
which preserve the reflection amplitude while changing the potential landscape. 
They correspond to a coordinate change $z\to\bz$ and rescaling $\psi\to\bpsi=\sqrt{\bz'(z)}\psi$ 
of the wave-function
We choose here a specific Liouville gauge discussed in \cite{Dufour2015jpb}, 
with the new coordinate $\bz = \phi_\dB/\varkappa$ proportional to the WKB phase 
$\phi_\dB$ ($\varkappa$ an arbitrary constant).
The latter is the integral $\phi_\dB=\int k_\dB dz$ of the WKB wave-vector $k_\dB =\sqrt{2m(E-V)}/\hbar$.
The Liouville transformation preserves the form of the Schr\"odinger equation 
with modified energy $\bE$ and potential  $\bV$ \cite{Dufour2015jpb}
\begin{eqnarray}
&&\bV(\bz)=\bE\,Q(z)\quad,\quad \bE=\varkappa^2 ~, \\
&&Q(z)= - \alpha_\dB^3 \frac{d^2 \alpha_\dB}{dz^2}\quad,\quad
\alpha_\dB\equiv \frac1{\sqrt{k_\dB}} \notag ~.
\end{eqnarray}
$Q(z)$ is the \textit{badlands} function  \cite{Berry1972,Friedrich2004a}
which marks the zones where the WKB approximation breaks down, that is also
where significant quantum reflection occurs \cite{Dufour2013qrefl}.

\begin{figure}[htb]
   \center
   \includegraphics[scale=0.4]{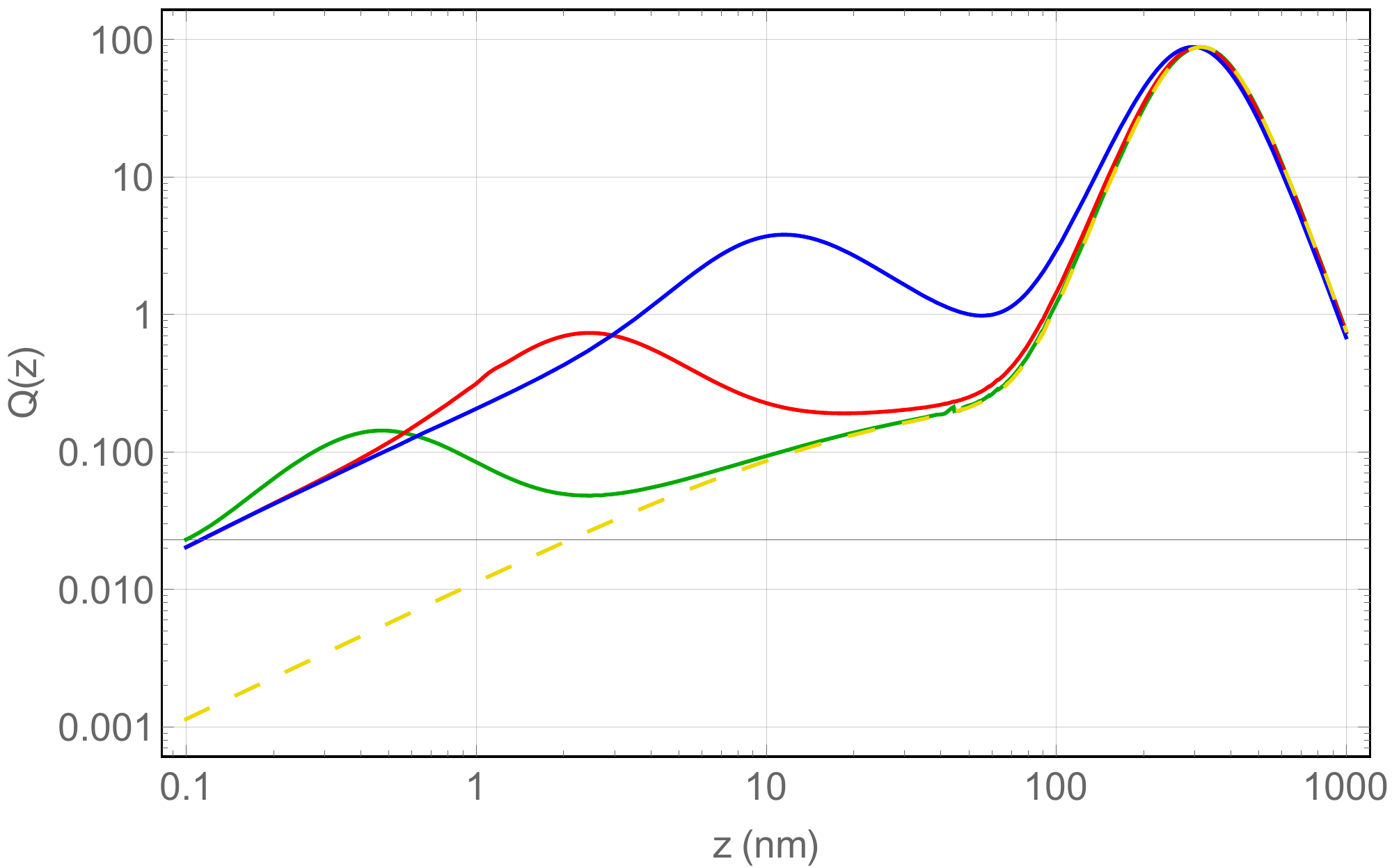}
   \caption{Badlands functions $Q(z)$ ($z$ in nm) calculated for an antihydrogen atom falling from the energy of the first quantum gravitational state onto a liquid helium film above a gold substrate. The three full lines correspond to three thicknesses of the film, with the same color code as for the points emphasized
in Fig.~\ref{quadrature}~: From bottom to top, the thickness of the film is 1nm (green), 5nm (red) and 20nm (blue). 
The dashed (yellow) curve corresponds to the naked substrate.}
   \label{CPLiouville}
\end{figure}

We draw in Fig.~\ref{CPLiouville} the {badlands} function $Q(z)$ for four different 
thicknesses of the film supported by a gold substrate. 
Three thicknesses correspond to the colored points emphasized in Fig.~\ref{a} 
for $d=$1nm (green), 5nm (red) and 20nm (blue). 
For the purpose of comparison with the case of a naked substrate,
a fourth plot is drawn for $d=$0 (dashed yellow curve). 
The plots for non null thicknesses show two peaks while only one peak 
appears for the plot of the naked substrate as well as for the plot
of a liquid helium bulk. 
The peak lying far from the surface is roughly the same for all curves,
and it is the same as for the naked substrate. The other one corresponds
to the approach to the helium film and its position depends on the 
thickness of the film. 

For films with non null thicknesses, 
the two peaks form a cavity where the matter wave can be stored. 
In the case studied here, the mirror closer to the material surface has a poorer reflectivity than the mirror
farther from the material surface. 
The presence of the cavity leads to a faster annihilation when atoms are trapped, which degrades 
the lifetime of the antihydrogen atom, as observed in Fig.~\ref{a}-\ref{tau}.
The interferences taking place in the cavity explain the oscillation patterns highlighted in Fig.~\ref{a} and \ref{quadrature}. The associated phase is related to the round-trip dephasing in the cavity, which is determined 
by the displacement to the left of the weaker peak and the change of the shape of the potential inside the cavity.
This discussion can be considered as a qualitative interpretation of the full calculations presented in the 
preceding section.

\section{Conclusion}

In this letter we have found theoretically a high reflection probability for antihydrogen atoms falling down 
onto thick enough liquid helium films.
We also predicted the presence of oscillations of the scattering
length as a function of thickness for liquid helium films supported by a substrate.
We interpreted the associated interference pattern as a consequence of the existence of two 
separated zones where significant reflection occurs. 

We have considered that low-temperature reflection properties of antihydrogen atoms from liquid helium films are essentially determined by the long-range part of potential, as it is known theoretically \cite{Carraro1992} and proven experimentally \cite{Yu1993} for the case of hydrogen-liquid helium interaction. 
Of course, it would be interesting to go further in the analysis by considering explicitly the effect of the 
short-range part of the interaction potential. 
This idea has been studied for the case of antihydrogen atom-helium atom interaction, 
and short-range repulsion predicted \cite{Strasburger2002,Jonsell2004,Strasburger2005,Froelich2012,Jonsell2012}. 
It would be worth performing similar calculations for the case of antihydrogen atom-liquid helium interaction 
studied in the present letter. These calculations may change the precise value of the lifetimes obtained above, but
they should not affect qualitatively our main statement, namely that long lifetimes are obtained for 
antimatter above a liquid helium bulk, which have very interesting applications
for new spectroscopic tests of the equivalence principle for antihydrogen atoms.

\smallskip

\textit{Acknowledgements -} 
We are grateful to our colleagues from GBAR and GRANIT collaborations 
for useful discussions.
We thank the referee for stimulating comments.

\end{document}